\documentclass[%
 reprint,
 amsmath,amssymb,
 aps,
 prstab,
]{revtex4-1}

\usepackage{dcolumn}	
\usepackage{bm}			
\usepackage{graphicx}	
\usepackage[nodayofweek]{datetime}	

\begin{document}

\title{Phase Jump Method for Efficiency Enhancement in Free-Electron Lasers}

\author{Alan Mak}
\email{alan.mak@maxiv.lu.se}
\author{Francesca Curbis}
\author{Sverker Werin}

\affiliation{MAX IV Laboratory, Lund University, P.O.~Box 118, SE-22100 Lund, Sweden}
\date{\today}

\begin{abstract}
The efficiency of a free-electron laser can be enhanced by sustaining the growth of the radiation power beyond the initial saturation. One notable method is undulator tapering, which involves the variation of the gap height and/or the period along the undulator. Another method is the introduction of phase jumps, using phase-shifting chicanes in the drift sections separating the undulator segments. In this article, we develop a physics model of this phase jump method, and verify it with numerical simulations. The model elucidates the energy extraction process in the longitudinal phase space. The main ingredient is the microbunch deceleration cycle, which enables the microbunched electron beam to decelerate and radiate coherently beyond the initial saturation. The ponderomotive bucket is stationary, and energy can even be extracted from electrons outside the bucket. The model addresses the selection criteria for the phase jump values, and the requirement on the undulator segment length. It also describes the mechanism of the final saturation. In addition, we discuss the similarities and differences between the phase jump method and undulator tapering, by comparing our phase jump model to the classic Kroll-Morton-Rosenbluth model.
\end{abstract}

\maketitle

\section{Introduction}

In most single-pass free-electron laser (FEL) facilities, the undulator line is segmented by drift sections, where instruments for beam focusing, trajectory correction and diagnostics are installed. Often time, phase shifters are also installed in the drift sections~\cite{EuropeanXFEL, FERMI, PAL_XFEL, Tischer}.

Conventional phase shifters are compact magnetic chicanes, made up of either permanent magnets or electromagnets. The magnetic chicane enables the increase of the electron path length in the drift section, thus adjusting the phase angle between the electron beam and the optical field.

A common use of phase shifters is phase correction, i.e.\ to make the phase angle in the preceding undulator segment equal to that in the subsequent undulator segment. This facilitates the constructive interference of the optical fields emitted in the two undulator segments. Without the phase correction, the velocity difference between electrons and light in the drift section can lead to an unwanted change in the phase angle, which causes destructive interference of the optical fields.

Another use of phase shifters is the suppression of the fundamental wavelength in harmonic lasing, so as to increase the spectral brightness of the desired harmonic~\cite{Harmonic_lasing_Brits, Harmonic_lasing_Russians}.

In addition to phase correction and fundamental wavelength suppression, phase shifters can also be used for sustaining the energy extraction beyond the initial saturation point, thus enhancing the FEL efficiency. This is achieved by applying appropriate phase jumps, purposely altering the phase angle between the electron beam and the optical field. The efficiency enhancement by this phase jump method is demonstrated numerically in Refs.~\cite{Varfolomeev} and~\cite{Ratner}.

In this article, we further the studies by developing a physics model of the phase jump method, and verifying it with numerical simulations. Our model illustrates the particle dynamics in the longitudinal phase space, and enables a deeper understanding of the physics behind the energy extraction process.

Apart from the phase jump method, another common technique for FEL efficiency enhancement is undulator tapering, which involves the variation of the undulator parameter and/or the undulator period along the undulator line~\cite{KMR, Alan2015}. In Ref.~\cite{Ratner}, the authors highlight the similarities between phase jumps and undulator tapering. In this article, we supplement the discussion by contrasting the differences between the two techniques, with the aid of our physics model.

The phase jump method is particularly useful in FEL facilities with fixed-gap undulators, which cannot be tapered easily. In such facilities, phase shifters can be installed in the existing drift sections, and the drift sections  can be expanded if necessary. With the phase jump method, the FEL efficiency can be enhanced without replacing all the fixed-gap undulator segments by variable-gap ones.

\section{Physics Model}

\subsection{Problem description}

Consider an FEL which comprises planar undulator segments, with a phase shifter installed in every drift section. All undulator segments have the same length $L_{\rm{segm}}$, meaning that the distance between every two successive phase shifters is constant throughout the FEL.

In the following, we develop a one-dimension, steady-state model of the phase jump method, which sustains the growth of radiation power at the fundamental wavelength beyond the initial saturation point.

In particular, we are interested in the effect of the phase jumps alone, in the absence of undulator tapering. We therefore restrict ourselves to a constant undulator period $\lambda_u = 2\pi/k_u$ and a constant undulator parameter
\begin{equation}
K = \frac{e B_0}{m_e c k_u},
\end{equation}
where $e$ is the absolute value of the electron charge, $m_e$ is the electron rest mass, $c$ is the speed of light, and $B_0$ is the peak undulator field.

In addition, we assume the amplitude $E_0$ and phase $\phi$ of the optical field to be slowly varying in the course of the FEL interaction.

\subsection{Energy definitions}

The energy of an electron can be expressed as $\gamma m_e c^2$, and the resonant energy $\gamma_R m_e c^2$ is defined by
\begin{equation}
\label{Eq:resonance}
\gamma_R = \sqrt{ \frac{\lambda_u}{2 \lambda} \left( 1+ \frac{K^2}{2} \right) },
\end{equation}
where $\lambda = 2\pi/k$ is the radiation wavelength.

In the case of monochromatic seeding, $\lambda$ is the wavelength of the seed radiation, and $\gamma_R$ is determined by $\lambda$ through Eq.~(\ref{Eq:resonance}). In the case of self-amplified spontaneous emission (SASE), $\gamma_R$ is determined by the energy of the incoming electron beam, and $\lambda$ is determined by $\gamma_R$ through Eq.~(\ref{Eq:resonance}).

With undulator tapering, the undulator parameter $K$ decreases with the distance $z$ along the FEL. According to the definition in Eq.~(\ref{Eq:resonance}), $\gamma_R$ decreases with $z$ to retain the radiation wavelength $\lambda$. In the absence of undulator tapering, however, $K$ is constant. According to the definition in Eq.~(\ref{Eq:resonance}), $\gamma_R$ remains constant to retain $\lambda$.

In other words, the resonant energy is constant in the phase jump method. This allows us to express the energy of an electron as the relative deviation from the resonant energy, by the variable
\begin{equation}
\eta \equiv \frac{\gamma - \gamma_R}{\gamma_R}.
\end{equation}
Even though the resonant energy is constant by definition, if the electrons themselves can decrease in $\eta$ beyond the initial saturation point, they can continue to transfer energy to the optical field.

\subsection{Phase definitions}

In the analysis of phase jumps, we are not interested in the absolute phase of an electron. Instead, it is more convenient to consider the phase of an electron with respect to the ponderomotive potential well, given by
\begin{equation}
\psi \equiv (k + k_u) z - c k t + \phi.
\end{equation}

Let $\psi_{\rm{orig}} \in [-\pi,\pi]$ be the original phase of an electron at the start point of a drift section, and $\psi_{\rm{targ}} \in [-\pi,\pi]$ be the target phase at the end point of the same drift section. The phase jump is then the difference:
\begin{equation}
\psi_{\rm{jump}} = \psi_{\rm{targ}} - \psi_{\rm{orig}}  \in [-2\pi,2\pi].
\end{equation}

Note the sign convention that a \textit{positive} $\psi_{\rm{jump}}$ corresponds to shifting the electron \textit{forward} in $\psi$. Also, $\psi_{\rm{jump}} = 0$ corresponds to mere phase correction, whereby the phase angle introduced by the electron-light velocity difference in the drift section is compensated exactly.

In practice, a conventional phase shifter applies the phase jump by increasing the electron path length. It can only shift electrons backward in $\psi$, but not forward. Thus, if the required $\psi_{\rm{jump}}$ is positive, we need to shift the electron backward to another potential well by a phase of $2 \pi n - \psi_{\rm{jump}}$, where $n$ is a positive integer.

As a side note, this phase jump method for efficiency enhancement can, in principle, be implemented in combination with the iSASE technique~\cite{iSASE} for bandwidth reduction. This is done by choosing a large $n$, so that the optical field emitted by the electrons towards the tail of the bunch may develop correlations with the electrons towards the head of the bunch, thus increasing the coherence length.

\subsection{Equations of motion}

In our model of the phase jump method, we concern ourselves with the electron dynamics in the longitudinal phase space $(\psi, \eta)$. In the undulator segments, the longitudinal dynamics of an electron can be described by two equations of motion:
\begin{eqnarray}
 \label{Eq:eta} \frac{d \eta}{dz} & = & - \frac{\Omega^2}{2 k_u} \sin \psi, \\
 \label{Eq:psi} \frac{d \psi}{dz} & = & 2 k_u \eta. 
\end{eqnarray}
A derivation of these equations is given in Ref.~\cite{Schmueser}.

In Eq.~(\ref{Eq:eta}), the quantity
\begin{equation}
\label{Eq:SyncFreq}
\Omega = \sqrt{\frac{e}{m_e c^2} \frac{k_u K f_B E_0}{\gamma_R^2}} \propto \sqrt{E_0}
\end{equation}
is the angular frequency of the synchrotron oscillation, which has the dimension of inverse length. Meanwhile, $f_B = J_0 (\xi) - J_1 (\xi)$ is the Bessel factor for planar undulators, with $\xi = K^2 / [2(K^2+2)]$. By substituting Eq.~(\ref{Eq:eta}) into the derivative of Eq.~(\ref{Eq:psi}), we can verify that the longitudinal dynamics satisfies the pendulum equation
\begin{equation}
\frac{d^2 \psi}{dz^2} + \Omega^2 \sin\psi = 0.
\end{equation}

\subsection{Phase space trajectories}

\begin{figure}[tb]
   \centering
   \includegraphics*[width=\columnwidth]{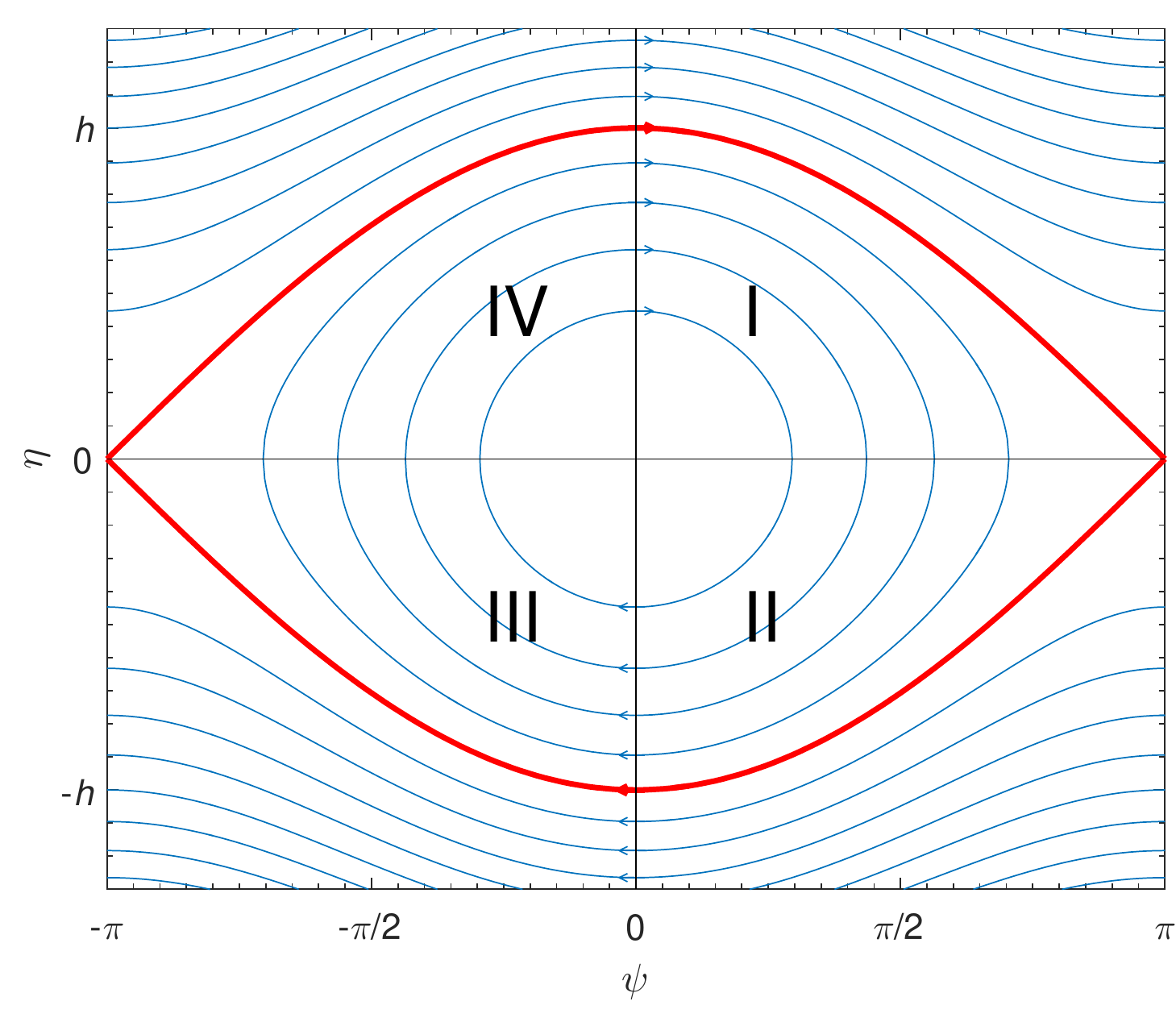}
   \caption{The longitudinal phase space $(\psi, \eta)$, with electron trajectories shown by the blue curves. The red curve is the \textit{separatrix}, and the region enclosed by it is the \textit{ponderomotive bucket}. The straight lines $\eta = 0$ and  $\psi = 0$ divide the space into four quadrants, as indicated by the Roman numerals.}
   \label{Fig:four_quadrants}
\end{figure}

The equations of motion~(\ref{Eq:eta}) and~(\ref{Eq:psi}) satisfy the Hamilton equations for the Hamiltonian
\begin{equation}
\label{Eq:H}
H (\psi, \eta) = c k_u \eta^2 + \frac{c \Omega^2}{2 k_u} (1- \cos\psi).
\end{equation}
The electron trajectories in the longitudinal phase space $(\psi, \eta)$ are given by the level set of the function $H (\psi, \eta)$, and are shown in Fig.~\ref{Fig:four_quadrants}.

In particular, the trajectory highlighted in red is known as the \textit{separatrix}. Along the separatrix, the Hamiltonian has the value
\begin{equation}
\label{Eq:H_sep}
H_{\rm{sep}} = H (\pm \pi,0) = \frac{c \Omega^2}{k_u}.
\end{equation}
The region enclosed by the separatrix is known as the \textit{ponderomotive bucket}. Within the bucket, $H < H_{\rm{sep}}$. The trajectories are closed orbits, and the electrons are \textit{trapped}. Outside the bucket, $H > H_{\rm{sep}}$. The trajectories are unbounded, and the electrons are \textit{detrapped}.

The maximum $\eta$ value along the separatrix gives the half-height of the bucket,
\begin{equation}
h = \frac{\Omega}{k_u}.
\end{equation}
Invoking the definition of the synchrotron frequency in Eq.~(\ref{Eq:SyncFreq}), we obtain the proportionality 
\begin{equation}
\label{Eq:h_propto_E0}
h \propto \sqrt{E_0},
\end{equation}
meaning that the bucket half-height $h$ increases with the optical field amplitude $E_0$.

In Fig.~\ref{Fig:four_quadrants}, the horizontal line $\eta = 0$ and the vertical line $\psi = 0$ divide the longitudinal phase space into four quadrants, as indicated by the Roman numerals.

In quadrants I and II, $\psi > 0$. In quadrants III and IV, $\psi < 0$. According to Eq.~(\ref{Eq:eta}), this implies $d\eta/ dz < 0$ in quadrants I and II, and $d\eta/ dz > 0$ in quadrants III and IV. In other words, electrons decelerate in quadrants I and II, and accelerate in quadrants III and IV. Due to the conservation of energy, energy is transferred to the optical field in quadrants I and II, and energy is absorbed from the optical field in quadrants III and IV.

In quadrants I and IV, $\eta > 0$. In quadrants II and III, $\eta < 0$. According to Eq.~(\ref{Eq:psi}), this implies that electrons have increasing $\psi$ in quadrants I and IV, and decreasing $\psi$ in quadrants II and III.

\subsection{Phase jump commencement}

The essence of the phase jump method is microbunch deceleration. The aim is to decelerate the microbunched beam after the initial saturation, so that it can continue to radiate coherently. Thus, the phase jumps should commence in the vicinity of the initial saturation point, where the microbunching is fully developed.

In the exponential regime, the phase shifters should be configured for $\psi_{\rm{jump}} = 0$, or there will be disruption in the microbunching development. For SASE FELs in particular, applying phase jumps in the exponential regime can also lead to a red or blue shift in the radiation wavelength, depending on the positions and magnitudes of the phase jumps~\cite{Harmonic_lasing_Russians}.

\subsection{Microbunch deceleration mechanism}

\begin{figure}[tb]
   \centering
   \includegraphics*[width=\columnwidth]{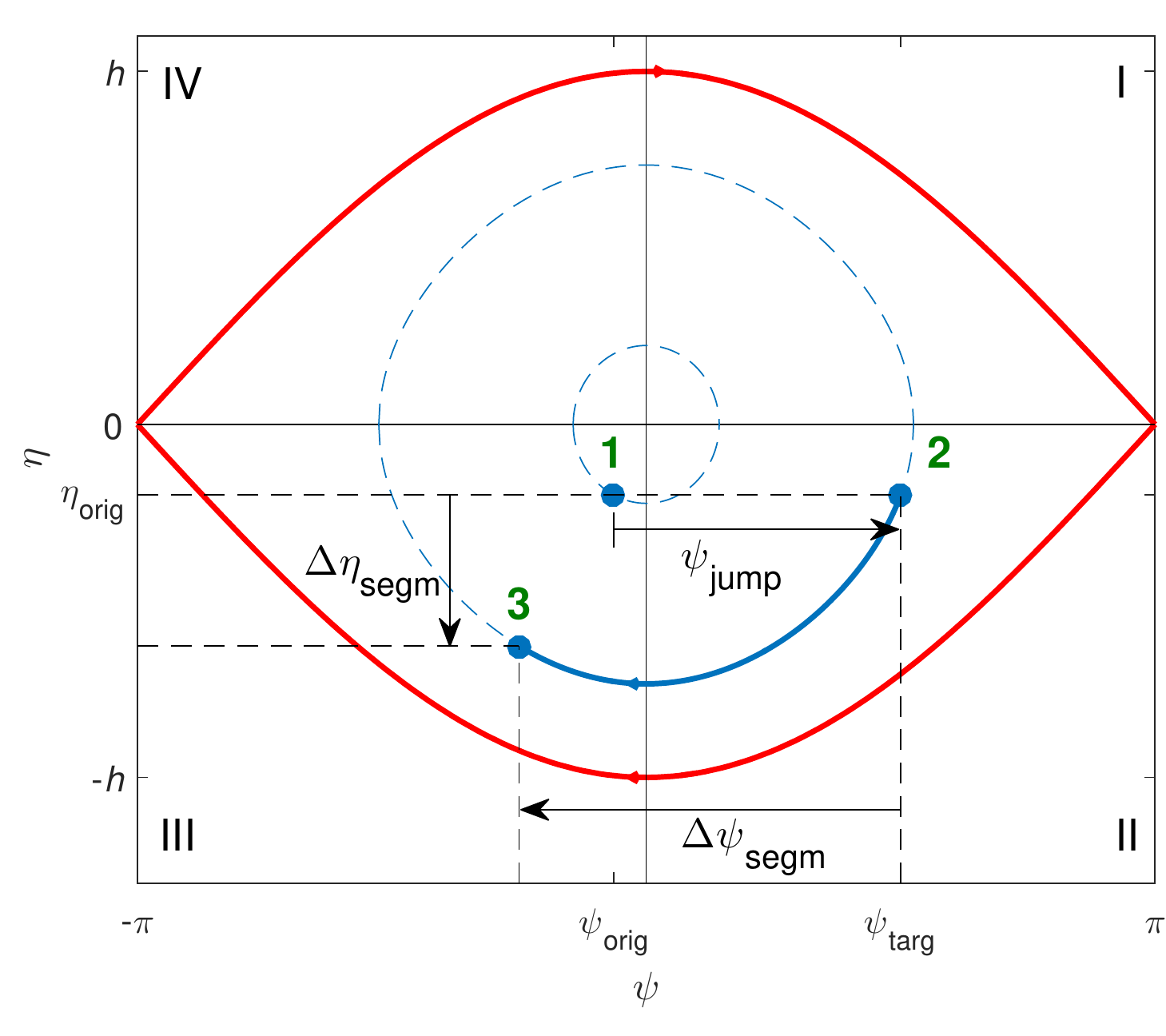}
   \caption{The microbunch deceleration cycle as illustrated by the movement of the average particle within the ponderomotive bucket. Position 1 corresponds to the start point of a drift section, position 2 the end point of the same drift section, and position 3 the end point of the subsequent undulator segment.}
   \label{Fig:decel_cycle}
\end{figure}

To analyze the microbunch deceleration, it is convenient to follow the motion of an average particle within the microbunch $\mu$. Let $(\bar{\psi}, \bar{\eta})$ be the coordinates of the average particle in the longitudinal phase space. They can be defined as
\begin{eqnarray}
\bar{\eta} & \equiv & \left\langle \eta \right\rangle_\mu =
\frac{1}{N} \sum\limits_{j \in \mu}^{N} \eta_j,  \label{Eq:eta_bar} \\
\bar{\psi} & \equiv & \arg \left\langle e^{-i\psi} \right\rangle_\mu =
-i \ln \left( \frac{1}{N} \sum\limits_{j \in \mu}^{N} e^{-i\psi_j} \right), \label{Eq:psi_bar}
\end{eqnarray}
where $N$ is the number of particles in $\mu$.

Microbunch deceleration takes place in quadrants I and II, where particles decelerate and transfer energy to the optical field (see Fig.~\ref{Fig:four_quadrants}). Thus, the general principle of the phase jump method is to maximize the time that the average particle spends in the deceleration quadrants (I and II), and minimize the time that the average particle spends in the acceleration quadrants (III and IV).

In the longitudinal phase space $(\psi, \eta)$, a phase jump moves a particle horizontally. If the average particle lies in quadrant IV, then we should choose a phase jump that moves it into quadrant I. If it lies in quadrant III, then we should move it into quadrant II. If it lies in quadrant II but is about to enter quadrant III, then we should move it to a slightly larger phase within quadrant II (i.e. away from quadrant III).

The mechanism is illustrated in Fig.~\ref{Fig:decel_cycle}. Suppose that the average particle has an original phase $\bar{\psi} = \psi_{\rm{orig}} <~0$ at the start point of a drift section, as indicated by position 1. We then apply a phase jump $\psi_{\rm{jump}} >~0$, so that the average particle arrives at a target phase $\bar{\psi} = \psi_{\rm{targ}} > 0$ at the end point of the drift section, as indicated by position 2.

At position 2, the electrons enter an undulator segment, and follow the phase space trajectories described by Eqs.~(\ref{Eq:eta}) and~(\ref{Eq:psi}). In particular, the average particle follows the blue solid curve. So long as $\bar{\psi} >~0$, the microbunch decelerates, and transfer energy to the optical field. During this energy transfer, the optical field amplitude $E_0$ increases slowly, and the bucket half-height $h$ increases slowly according to the proportionality~(\ref{Eq:h_propto_E0}).

The average particle then arrives at position 3, as it reaches the end point of the undulator segment. Depending on the length $L_{\rm{segm}}$ of the undulator segment, position 3 can be located in either quadrant II or III. In quadrant III, where $\bar{\psi} <~0$, the microbunch absorbs energy from the optical field.

Within the undulator segment, $\bar{\psi}$ and $\bar{\eta}$ of the average particle have changed by $\Delta\psi_{\rm{segm}}$ and $\Delta\eta_{\rm{segm}}$, respectively. Provided that $\Delta\eta_{\rm{segm}} <~0$, the microbunch has a net deceleration, and hence a net energy transfer to the optical field.

The end point of the undulator segment is also the start point of the next drift section. We can then repeat this microbunch deceleration cycle, by taking position 3 of the old cycle as position 1 of the new cycle. The cycle can continue until the end of the last undulator segment.

As the cycle continues, the microbunch moves towards the bottom of the ponderomotive bucket. Close to the bottom of the bucket, further phase jumps will move the microbunch out of the bucket. In other words, we can divide the energy extraction process beyond the initial saturation point into three main stages: (i) the in-bucket regime, (ii) the out-of-bucket regime, and (iii) the final saturation regime.

With an appropriate choice of the target phase $\psi_{\rm{targ}}$ in every phase jump, we can have $\Delta\eta_{\rm{segm}} <~0$ in every undulator segment between the initial saturation and the final saturation. Obtaining the precise value of the optimal $\psi_{\rm{targ}}$ is a matter of empirical phase scan. But from a theoretical perspective, there are general criteria for a good choice of $\psi_{\rm{targ}}$ within the deceleration quadrants.

\subsection{In-bucket regime}

The in-bucket regime is the first stage beyond the initial saturation point. At this stage, microbunch deceleration takes place along the closed orbits within the ponderomotive bucket.

\subsubsection{Lower bound for good target phase}
\label{Sec:psi_min}

In the single-cycle microbunch deceleration illustrated in Fig.~\ref{Fig:decel_cycle}, the energy extraction is the most efficient if the average particle stays within the deceleration quadrants throughout the entire undulator segment, and never manages to enter quadrant III. For this to be the case, the chosen target phase $\psi_{\rm{targ}}$ must satisfy the criterion
\begin{eqnarray}
\nonumber \psi_{\rm{targ}} - | \Delta \psi_{\rm{segm}} |& \geq & 0 \\
\Leftrightarrow \psi_{\rm{targ}} & \geq & | \Delta \psi_{\rm{segm}} |. \label{Eq:psi_min_criterion}
\end{eqnarray}

In order to proceed from here, we obtain an expression for $\Delta \psi_{\rm{segm}}$ by integrating both sides of Eq.~(\ref{Eq:psi}) with respect to $z$ over one undulator segment. This yields
\begin{equation}
| \Delta \psi_{\rm{segm}}| =  2 k_u \left| \int_{z'}^{z'+L_{\rm{segm}}}\bar{\eta}(z) dz \right|.
\end{equation}
Within the undulator segment, we expect the average particle to decelerate, and $\bar{\eta}(z)$ should therefore be more negative than the original value $\eta_{\rm{orig}}$ right before the undulator segment. Hence,
\begin{equation}
\label{Eq:Delta_psi_segm}
| \Delta \psi_{\rm{segm}} | \geq  2 k_u |\eta_{\rm{orig}}| L_{\rm{segm}}.
\end{equation}

Combining the inequalities~(\ref{Eq:psi_min_criterion}) and~(\ref{Eq:Delta_psi_segm}), we obtain the lower bound $\psi_{\rm{min}}$ for the choice of $\psi_{\rm{targ}}$:
\begin{equation}
\label{Eq:psi_min}
\psi_{\rm{targ}} \geq 2 k_u |\eta_{\rm{orig}}| L_{\rm{segm}} \equiv \psi_{\rm{min}}.
\end{equation}
When configuring each phase shifter, $\psi_{\rm{targ}}$ needs to be at least $\psi_{\rm{min}}$, for the average particle to \textit{have a chance} of avoiding the acceleration quadrants. If $\psi_{\rm{targ}}$ is less than $\psi_{\rm{min}}$, then the average particle will \textit{definitely} enter quadrant III within the upcoming undulator segment.

\subsubsection{Upper bound for good target phase}

During the in-bucket regime, we should keep the average particle in the bucket as long as possible. This allows us to fully exploit the in-bucket regime before the average particle becomes detrapped.

Thus, the upper bound $\psi_{\rm{max}}$ for the target phase $\psi_{\rm{targ}}$ is given by the separatrix in the deceleration quadrants (see Fig.~\ref{Fig:decel_cycle}). This can be expressed mathematically as
\begin{equation}
\label{Eq:psi_max_criterion}
H (\psi_{\rm{max}}, \eta_{\rm{orig}}) = H_{\rm{sep}}.
\end{equation}

To proceed from Eq.~(\ref{Eq:psi_max_criterion}), we can substitute the right-hand side by Eq.~(\ref{Eq:H_sep}), and the left-hand side by Eq.~(\ref{Eq:H}) with $(\psi,\eta) = (\psi_{\rm{max}}, \eta_{\rm{orig}})$. Cognizant of the fact that $0 \leq \psi_{\rm{max}} \leq \pi$, we can then solve for $\psi_{\rm{max}}$, and obtain the expression
\begin{equation}
\label{Eq:psi_max}
\psi_{\rm{max}} = 2 \arccos \left( \frac{k_u^2 |\eta_{\rm{orig}}|}{\Omega^2} \right).
\end{equation}
This is the upper bound for the choice of $\psi_{\rm{targ}}$, in order to avoid entering the out-of-bucket regime.

\subsubsection{Undulator segment length}

Within the first few cycles of the mechanism depicted in Fig.~\ref{Fig:decel_cycle}, the average particle should have reached the $\eta < 0$ region, i.e. quadrants II and III. For the microbunch deceleration to be efficient, the average particle should stay within quadrant II, without entering quadrant III. However, this will not be possible if $\Delta \psi_{\rm{segm}}$ is too large (see Fig.~\ref{Fig:decel_cycle}). From the inequality~(\ref{Eq:Delta_psi_segm}), we notice that $\Delta \psi_{\rm{segm}}$ increases with the undulator segment length $L_{\rm{segm}}$. This imposes an upper limit on $L_{\rm{segm}}$. 

As a particle undergoes one complete orbit in the ponderomotive bucket, it travels down the undulator magnet by a distance of one synchrotron period $L_{\rm{sync}} = 2 \pi/ \Omega$. As the particle sweeps across one quadrant in the bucket, it undergoes a quarter of a complete orbit, and travels down the undulator magnet by a distance of $L_{\rm{sync}}/4$.

Thus, for the average particle to stay within a single quadrant (namely, quadrant II), an undulator segment should be no longer than $L_{\rm{sync}}/4$. Since $L_{\rm{sync}}$ varies with $z$, the requirement for the undulator segment length is
\begin{equation}
\label{Eq:Lsegm_required}
L_{\rm{segm}} < \frac{1}{4} \min [L_{\rm{sync}}(z)].
\end{equation}

\subsection{Regime transition}

As the microbunch deceleration cycle continues, the relative energy deviation $\bar{\eta}$ of the average particle becomes more and more negative, meaning that $|\eta_{\rm{orig}}|$ becomes larger with every phase jump.

Throughout the in-bucket regime, $\psi_{\rm{min}}$ increases with $|\eta_{\rm{orig}}|$ according to Eq.~(\ref{Eq:psi_min}), and $\psi_{\rm{max}}$ decreases with $|\eta_{\rm{orig}}|$ according to Eq.~(\ref{Eq:psi_max}). As the average particle is close to the bottom of the bucket, we will eventually encounter a scenario where $\psi_{\rm{min}} > \psi_{\rm{max}}$.

In such a scenario, it is no longer possible to choose a target phase $\psi_{\rm{targ}}$ in the range of $\psi_{\rm{min}} \leq \psi_{\rm{targ}} \leq \psi_{\rm{max}}$. We are then forced to choose $\psi_{\rm{targ}} > \psi_{\rm{max}}$, and move the average particle out of the bucket. This marks the end of the in-bucket regime, and the beginning of the out-of-bucket regime. 

\subsection{Out-of-bucket regime}

\begin{figure}[tb]
   \centering
   \includegraphics*[width=\columnwidth]{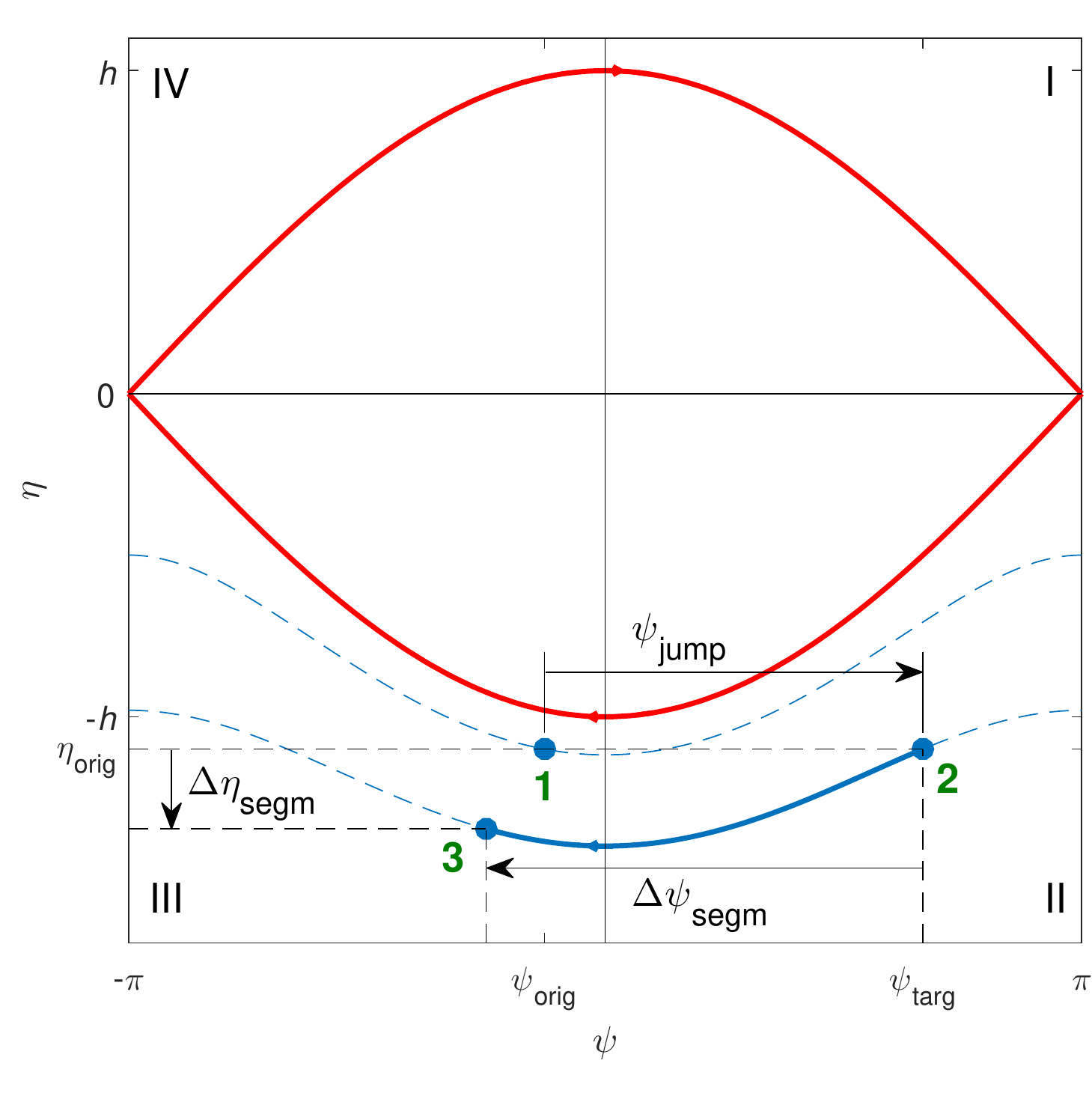}
   \caption{The microbunch deceleration cycle as illustrated by the movement of the average particle outside the bucket. Position 1 corresponds to the start point of a drift section, position 2 the end point of the same drift section, and position 3 the end point of the subsequent undulator segment.}
   \label{Fig:out-of-bucket_mechan}
\end{figure}

In the out-of-bucket regime, the trajectories in the longitudinal phase space $(\psi, \eta)$ are unbounded. Nonetheless, microbunch deceleration is possible. The mechanism is similar to that in the in-bucket regime, and is illustrated in Fig.~\ref{Fig:out-of-bucket_mechan}.

\subsubsection{Deceleration efficiency}

The deceleration efficiency in each undulator segment depends on the slope of the particle trajectory in the deceleration quadrants of the $(\psi, \eta)$ space. The steeper is the slope, the higher is the rate at which a particle loses energy.

The slope is given by the derivative $d \eta / d \psi$. Dividing Eq.~(\ref{Eq:eta}) by Eq.~(\ref{Eq:psi}), we can obtain an expression for the derivative as follows:
\begin{equation}
\label{Eq:decel_eff}
\left| \frac{d \eta}{d \psi} \right| =  \frac{\Omega^2}{4 k_u^2 |\eta|} \sin\psi.
\end{equation}
Note that in the deceleration quadrants, we have $0 \leq \psi \leq \pi$ and hence $0 \leq \sin\psi \leq 1$. As a result, $|d \eta / d \psi|$ is inversely proportional to $|\eta|$, and the deceleration efficiency decreases with $|\eta|$.

For the in-bucket regime, $|\eta| < h$. For the the out-of-bucket regime, $|\eta| > h$. Thus, the deceleration efficiency is lower in the out-of-bucket regime than in the in-bucket regime. As the microbunch deceleration cycle continues, the deceleration efficiency decreases with every undulator segment.

\subsubsection{Lower bound for good target phase}

For the in-bucket regime, the deceleration in an undulator segment can be made more efficient by keeping the average particle in the deceleration quadrants and preventing it from entering the acceleration quadrants. This argument also applies to the out-of-bucket regime. Thus, the lower bound for a good $\psi_{\rm{targ}}$ in the out-of-bucket regime is also given by Eq.~(\ref{Eq:psi_min}):
\begin{equation*}
\psi_{\rm{min}} = 2 k_u |\eta_{\rm{orig}}| L_{\rm{segm}}.
\end{equation*}

\subsubsection{Upper bound for good target phase}

In the out-of-bucket regime, the average particle is already outside the bucket. The separatrix does not impose any  limit on $\psi_{\rm{targ}}$. In principle, the upper limit of $\psi_{\rm{targ}}$ in the out-of-bucket regime is $\pi$, which is the maximum phase in the deceleration quadrants.

However, it is not favourable to let the average particle get too close to $\pi$, or a fraction of the particles within the microbunch will leak into the $\pi < \psi < 3\pi$ region, which corresponds to the acceleration quadrants associated with the bucket ahead. In that region, particles absorb energy from the optical field.

The precise upper bound for the choice of $\psi_{\rm{targ}}$ depends on the $\psi$ spread of the microbunch. But roughly speaking, the upper bound for a good $\psi_{\rm{targ}}$ is \textit{slightly below} $\pi$.

\subsection{Final saturation regime}

According to the relation (\ref{Eq:Delta_psi_segm}), $|\Delta \psi_{\rm{segm}}|$ increases with $|\eta_{\rm{orig}}|$. At some point in the out-of-bucket regime, $|\eta_{\rm{orig}}|$ will have become so large that
\begin{equation}
|\Delta \psi_{\rm{segm}}| = \pi.
\end{equation}
This signifies the onset of the final saturation regime.

Beyond that point, it is no longer possible to prevent the average particle from moving into quadrant III within a single undulator segment, regardless of the choice of $\psi_{\rm{targ}}$ (see Fig.~\ref{Fig:out-of-bucket_mechan}). The microbunch deceleration cycle then becomes inefficient, and $\Delta \eta_{\rm{segm}}$ approaches zero.

As $\Delta \eta_{\rm{segm}}$ approaches zero, the inequality~(\ref{Eq:Delta_psi_segm}) can be approximated by
\begin{equation}
| \Delta \psi_{\rm{segm}} | \approx 2 k_u |\eta_{\rm{orig}}| L_{\rm{segm}} \equiv \psi_{\rm{min}}.
\end{equation}
With this approximation, the relative energy deviation at the onset of the final saturation regime is then
\begin{equation}
|\eta_{\rm{orig}}| =  \frac{\pi}{ 2 k_u L_{\rm{segm}} }. \label{Eq:final_sat_onset}
\end{equation}

The final saturation point is reached when $\Delta \eta_{\rm{segm}} \geq~0$, i.e.~when it is no longer possible to maintain a net transfer of energy from microbunch to the optical field.

\subsection{Small subtlety about phase jump}

When specifying a phase jump, it is important to remember that the specified value is only valid for particles at a certain reference energy. In other words, a particle which is not at the reference energy will experience a different phase jump from the specified value.

In applying the phase jump method, it is convenient to use the resonant energy $\gamma_R m_e c^2$ as the reference energy, as it is constant. But in our physics model, we are mainly concerned about the phase jump applied to the average particle, which has $\gamma \neq \gamma_R$ in general. Therefore, we need a conversion formula between the phase jump $\psi_{\rm{jump}}^{\rm{A}}$ for the average particle and the phase jump $\psi_{\rm{jump}}^{\rm{R}}$ for particles at the resonant energy.

To obtain such a conversion formula, we model the phase-shifting chicane as a one-period undulator, with undulator period $\hat{\lambda_u}$ and deflection parameter $\hat{K}$. After the one undulator period, the slippage $\hat{\lambda}$ is given by
\begin{equation}
\label{Eq:phase_shifter_undulator}
\hat{\lambda} = \frac{\hat{\lambda_u}}{2 \gamma^2} \left( 1 + \frac{\hat{K}^2}{2} \right).
\end{equation}

For particles at the resonant energy, $\gamma = \gamma_R$. The slippage $\hat{\lambda}$ is related to the phase jump $\psi_{\rm{jump}}^{\rm{R}}$ by
\begin{equation}
\label{Eq:psi_jump_R}
\frac{\psi_{\rm{jump}}^{\rm{R}}}{2 \pi} = - \frac{\hat{\lambda}}{\lambda},
\end{equation}
where $\lambda$ (without the caret) is the actual radiation wavelength of the FEL. The negative sign in the equation arises from the sign convention of $\psi_{\rm{jump}}^{\rm{R}}$.

In general, the average particle has $\gamma \neq \gamma_R$. In the same one-period undulator, the average particle experiences a different slippage $\hat{\lambda} + \Delta \hat{\lambda}$, which is related to the phase jump $\psi_{\rm{jump}}^{\rm{A}}$ by
\begin{equation}
\label{Eq:psi_jump_A}
\frac{\psi_{\rm{jump}}^{\rm{A}}}{2 \pi} = - \frac{\hat{\lambda} + \Delta \hat{\lambda}}{\lambda},
\end{equation}

To proceed from here, we take the differential on both sides of Eq.~(\ref{Eq:phase_shifter_undulator}), and obtain
\begin{eqnarray}
\Delta \hat{\lambda} & = & \frac{\hat{\lambda_u}}{2} \left( 1 + \frac{\hat{K}^2}{2} \right) \left( \frac{-2 \Delta \gamma}{\gamma^3} \right) \nonumber \\
& = & - 2 \hat{\lambda} \frac{\Delta \gamma}{\gamma} = - 2  \hat{\lambda} \bar{\eta}. \label{Eq:slippage_diff}
\end{eqnarray}
As usual, $\bar{\eta}$ is the relative energy deviation of the average particle. Using Eqs.~(\ref{Eq:psi_jump_R}), (\ref{Eq:psi_jump_A}) and (\ref{Eq:slippage_diff}), we can eliminate $\lambda$, $\hat{\lambda}$ and $\Delta \hat{\lambda}$. This results in the conversion formula
\begin{equation}
\psi_{\rm{jump}}^{\rm{A}} = \psi_{\rm{jump}}^{\rm{R}} (1 + 2 \bar{\eta}).
\end{equation}

\subsection{Comparison with undulator tapering}

Apart from the phase jump method, undulator tapering is another common technique for efficiency enhancement in FELs. In this subsection, we compare and contrast the two techniques. In particular, we discuss the similarities and differences between our phase jump model and the classic Kroll-Morton-Rosenbluth (KMR) model~\cite{KMR} of undulator tapering.

Both models make use of one-dimensional Hamiltonian mechanics to describe the particle dynamics in the longitudinal phase space $(\psi, \eta)$. The KMR tapering model follows the motion of a resonant particle, which defines the stable point in the middle of the ponderomotive bucket. Our phase jump model follows the motion of the average particle within the microbunch.

In the KMR model, we directly control the undulator parameter $K$, and move the resonant particle vertically in the $(\psi, \eta)$ space. In our phase jump model, we directly control the phase jump $\psi_{\rm{jump}}$, and move the average particle horizontally in the $(\psi, \eta)$ space.

In both models, the energy extraction is sustained beyond the initial saturation point by bringing a fraction of particles towards lower energies. As these particles decelerate, energy is transferred to the optical field, due to the conservation energy.

However, the underlying principle of the particle deceleration is different in the two models. In the KMR model, particle deceleration relies on the deceleration of the bucket itself. In our phase jump model, particle deceleration relies on the microbunch deceleration cycle.

In the KMR model, the bucket moves towards lower energies during the energy extraction. In this process, the phase of the resonant particle increases, and the width of the bucket decreases.  In our phase jump model, the bucket is stationary, and does not reduce in width. During the energy extraction process, the optical field amplitude increases, and the height of the bucket increases.

In the KMR model, particles need to be trapped in the bucket in order to decelerate. In our phase jump model, the microbunch deceleration cycle continues in the out-of-bucket regime. Energy extraction outside the bucket is impossible for the former, but possible for the latter.

In the KMR model, the efficiency of particle deceleration is determined by the rate at which the bucket decreases in energy. This, in turn, depends on the $dK/dz$, the rate at which the undulator parameter decreases along the undulator line. In the phase jump model, the efficiency of particle deceleration is determined by $d\eta/d\psi$, the slope of the particle trajectory in the longitudinal phase space. This, in turn, depends on the relative energy deviation $\bar{\eta}$ of the average particle, as evident by Eq.~(\ref{Eq:decel_eff}).

In the KMR model, the undulator segment length required for the efficient deceleration of particles is $L_{\rm{segm}} < \min [L_{\rm{sync}}(z)]$, as discussed in Ref.~\cite{Alan2015}. In our phase jump model, the requirement is $L_{\rm{segm}} < \min [L_{\rm{sync}}(z)]/4$, which is a more stringent one.

In the KMR model, the main causes of the final saturation are the weakening of refractive guiding and the detrapping of particles. In our phase jump model, the main causes of the final saturation are the decrease of $|d\eta/d\psi|$ and increase of $|\Delta \psi_{\rm{segm}}|$ with particle energy.

\section{Numerical simulations}

\subsection{Case definition}

Our physics model of the phase jump method is a one-dimensional and steady-state one. For the purpose of verifying the model, we perform a three-dimensional and steady-state simulation study, using the numerical simulation code GENESIS~\cite{GENESIS}.

We first define a case for the simulation study. The main parameters are listed in Table~\ref{Table:sim_par}.

\begin{table}[htb]
\caption{Simulation parameters.}
\label{Table:sim_par}
\begin{ruledtabular}
\begin{tabular}{lcc}
\textrm{Parameter}					&\textrm{Symbol}				&\textrm{Value}\\
\colrule \\ [-1em]
Electron beam energy				& $\gamma m_e c^2$				& 5 GeV \\
Energy spread						& $\sigma_{\gamma} / \gamma$	& $1 \times 10^{-4}$ \\
Beam current						& $I$				    		& 3 kA \\
Normalized emittance        		& $\varepsilon_{x,y}$			& 0.4 mm mrad \\
Average of beta	function			& $\bar{\beta}_{x,y}$			& 7 m \\
Radiation wavelength				& $\lambda$ 					& 2 \AA \\
Undulator period					& $\lambda_{w}$					& 20 mm \\
Undulator parameter					& $K$							& 1.35 \\
Length of each undulator segment	& $L_{\rm{segm}}$				& 1 m \\
Length of each drift section		& $L_{\rm{drift}}$				& 0.2 m \\
\end{tabular}
\end{ruledtabular}
\end{table}

In the chosen case, SASE is the start-up mechanism of the FEL. The effective shot-noise power is 830~W.

Planar undulator segments are used. Undulator tapering is not implemented.

The lattice for strong focusing is in a FODO configuration, with one quadrupole magnet in every other drift section. The length of the FODO cell is 4.8~m. Within the FODO cell, the centres of the two quadrupole magnets are separated by a distance of 2.4~m. The length of each quadrupole magnet is 80~mm.

The strengths of the quadrupole magnets and the initial twiss parameters are matched self-consistently to give the average beta $\bar{\beta}_{x,y}$ specified in Table~\ref{Table:sim_par}.

In the GENESIS simulations, we place a phase shifter in every drift section, by putting an \texttt{AD} element in the external magnet file. We control the phase jump by setting the \texttt{AD} element to an appropriate value.

\subsection{Initial saturation}

In order to obtain information about the initial saturation point, we first run the simulation in the absence of phase jumps, by setting $\psi_{\rm{jump}} = 0$ for all drift sections.

The simulation shows that the initial saturation occurs at $z = 38.2$ m within the 32th undulator segment. The saturation power is 2 GW. At the initial saturation point, the bunching factor $b = | \langle e^{-i \psi} \rangle |$ is the highest, and has a value of 0.4.

\subsection{Phase jump commencement}

We then repeat the simulation with the introduction of phase jumps. We start the phase jumps in the vicinity of the initial saturation point, where the microbunching is fully developed. Thus, the first non-zero phase jump occurs in the drift section at $z = 37$~m, immediately preceding the 32th undulator segment. Meanwhile, we keep $\psi_{\rm{jump}} = 0$ for all the drift sections before $z = 37$~m.

While there are infinite possible sets of phase jump values, we shall discuss one chosen set which yields an increase in radiation power beyond the initial saturation. It is possible to obtain even higher radiation power by fine-tuning the phase jump values. However, the purpose of this simulation study is to verify the physics model, and not to optimize the radiation power by phase scanning.

\subsection{Radiation power evolution}

\begin{figure*}[tb]
   \centering
   \includegraphics*[width=0.95\textwidth]{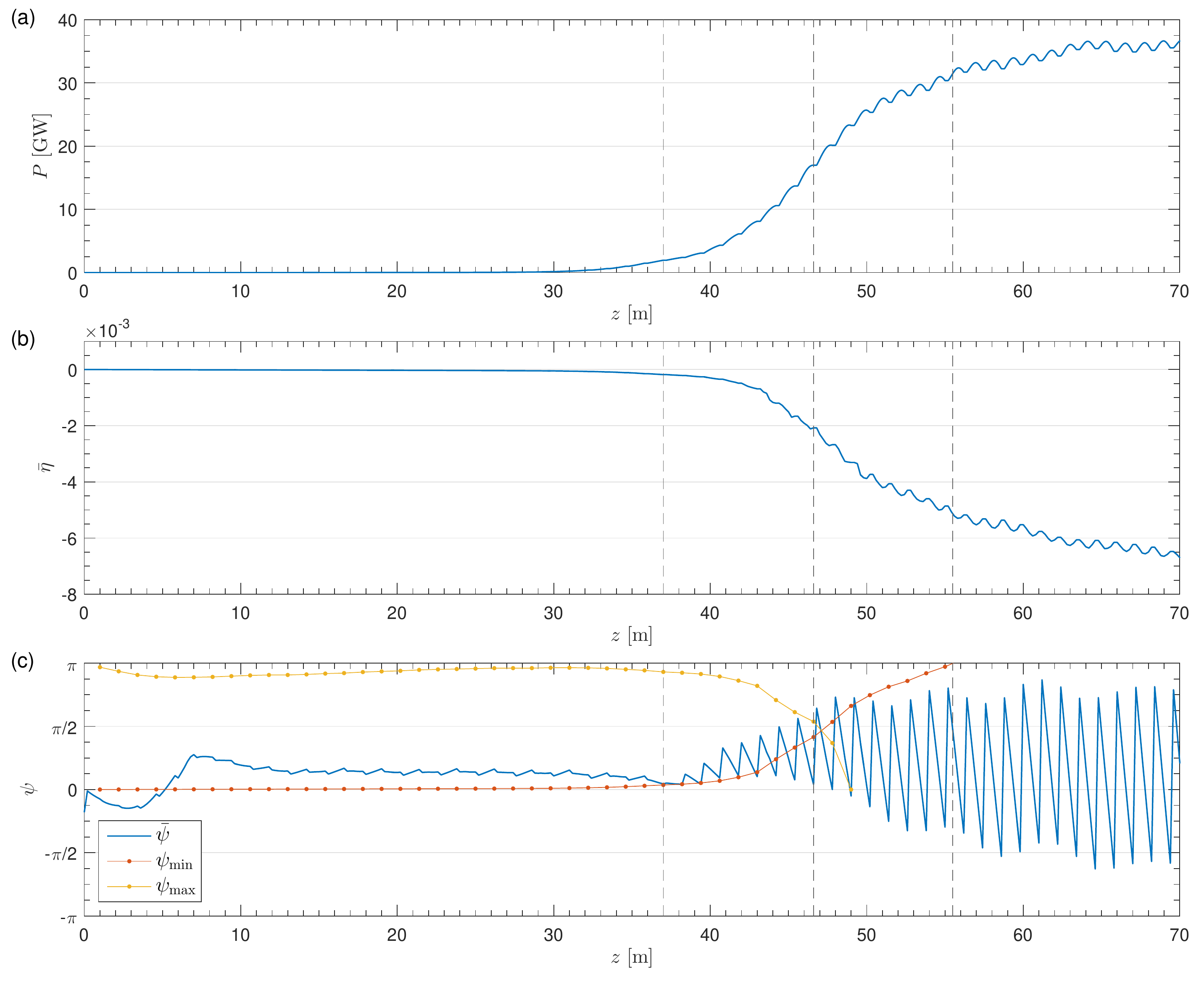}
   \caption{Simulation results. The following quantities are plotted as functions of the distance $z$ along the undulator line: (a) the radiation power, (b) the relative energy deviation $\bar{\eta}$ of the average particle, and (c) the ponderomotive phase $\bar{\psi}$ of the average particle (in blue), together with the lower bound $\psi_{\rm{min}}$ (in red) and the upper bound $\psi_{\rm{max}}$ (in yellow) for the target phase $\psi_{\rm{targ}}$ in each drift section. The dashed vertical lines mark the beginning of the in-bucket, out-of-bucket and final saturation regimes.}
   \label{Fig:evolution_curves}
\end{figure*}

Figure~\ref{Fig:evolution_curves}(a) shows the radiation power $P$ as a function of $z$, for the chosen set of phase jumps. Along the radiation power curve, there are short, straight, horizontal sections, where the power is neither increasing nor decreasing. These sections correspond to the drift sections, where the radiation does not exchange energy with the electron beam.

After the first non-zero phase jump at $z = 37$~m, the radiation power continues to grow. Up until $z = 50$~m, the power grows monotonically. But after that, the power fluctuates. Within every undulator segment, the power first increases, and then decreases. Nonetheless, there is still a net power increase.

In the undulator segment which begins at $z = 64.8$~m, there is no longer a net power increase within an undulator segment. This indicates the arrival of the final saturation. The power at the final saturation is 36 GW, which is 18 times the power at the initial saturation.

The FEL efficiency can be defined as the power ratio of the output radiation to the injected electron beam. With this definition, the chosen set of phase jumps enhance the FEL efficiency by a factor of 18. It is possible to obtain an even larger enhancement factor by fine-tuning the phase jump values. But again, the purpose of this simulation study is to verify the physics model, and not to perform a thorough optimization by phase scanning.

\subsection{Energy evolution of the average particle}

We now turn our attention to the average particle within the microbunch. The relative energy deviation $\bar{\eta}$ of the average particle is defined by Eq.~(\ref{Eq:eta_bar}). Figure~\ref{Fig:evolution_curves}(b) shows $\bar{\eta}$ as a function of $z$. 

Before the first non-zero phase jump at $z = 37$~m, $\bar{\eta}$ remains close to zero, meaning that the energy of the average particle is close to the resonant energy.

After the first non-zero phase jump at $z = 37$~m, $\bar{\eta}$ exhibits an overall decreasing trend. The energy of the average particle deviates further and further from the resonant energy. This is an evidence of microbunch deceleration. 

As $\bar{\eta}$ becomes more and more negative, the rate of $\bar{\eta}$ decrease becomes lower and lower. This agrees with the prediction of Eq.~(\ref{Eq:decel_eff}).

Up until $z = 50$~m, $\bar{\eta}$ decreases monotonically. But after that, $\bar{\eta}$ fluctuates. Comparing Fig.~\ref{Fig:evolution_curves}(a) and~(b), we notice that a decrease in $\bar{\eta}$ corresponds to an increase in $P$, and vice versa. This can be explained by the conservation of energy. When $\bar{\eta}$ decreases, the microbunch loses energy. This energy is transferred to the radiation, leading to an increase in $P$.

The onset of the final saturation regime is defined by Eq.~(\ref{Eq:final_sat_onset}). According to this definition,  the final saturation regime begins when $\bar{\eta} = 5 \times 10^{-3}$. As seen in Fig.~\ref{Fig:evolution_curves}(b), this corresponds to $z = 55.5$~m.

\subsection{Phase evolution of the average particle}

The ponderomotive phase $\bar{\psi}$ of the average particle is defined by Eq.~(\ref{Eq:eta_bar}). Figure~\ref{Fig:evolution_curves}(c) shows $\bar{\psi}$ as a function of $z$. After the first non-zero phase jump at $z = 37$~m, $\bar{\psi}$ oscillates in $z$. The upward slopes coincide with the drift sections, while the downward slopes coincide with the undulator segments. In other words, a crest coincides with the start point of an undulator segment, while a trough coincides with the end point of an undulator segment.

In terms of the microbunch deceleration cycle (see Figs.~\ref{Fig:decel_cycle} and~\ref{Fig:out-of-bucket_mechan}), a crest corresponds to position 2, while a trough corresponds to position 1 or 3. The period of the oscillation is one microbunch deceleration cycle. The crest value of each cycle is the target phase $\psi_{\rm{targ}}$. The phase change represented by an upward slope is $\psi_{\rm{jump}}$, while the phase change represented by a downward slope is $\Delta \psi_{\rm{segm}}$.

Figure~\ref{Fig:evolution_curves}(c) also shows the lower bound $\psi_{\rm{min}}$ and the upper bound $\psi_{\rm{max}}$ for the target phase $\psi_{\rm{targ}}$ in each drift section. The values are given by Eqs.~(\ref{Eq:psi_min}) and~(\ref{Eq:psi_max}). Recall that $\psi_{\rm{max}}$ is defined by the separatrix of the ponderomotive bucket in the $\psi \geq 0$ region, and that $\psi_{\rm{min}}$ is the minimum requirement for the average particle to avoid entering the $\psi < 0$ region.

In Fig.~\ref{Fig:evolution_curves}(c), the region immediately after the first non-zero phase jump at $z = 37$~m is the in-bucket regime, as evident by the fact that $\bar{\psi} < \psi_{\rm{max}}$. The decrease of $\psi_{\rm{max}}$ with $z$ reflects that the average particle is moving towards the bottom of the bucket.

In the in-bucket regime, $\psi_{\rm{targ}}$ is made to increase with $\psi_{\rm{min}}$, so as to fulfill the requirement that $\psi_{\rm{targ}} > \psi_{\rm{min}}$. As a result, the average particle is prevented from entering the $\psi < 0$ region. Within every undulator segment, the average particle transfers energy to the radiation, without absorbing energy from the radiation. This explains the monotonic decrease of $\bar{\eta}$ [see Fig.~\ref{Fig:evolution_curves}(b)].

Prior to $z = 46.5$~m, the choice of $\psi_{\rm{targ}}$ satisfies the requirement that $\psi_{\rm{min}} < \psi_{\rm{targ}} < \psi_{\rm{max}}$. But in the vicinity of $z = 46.5$~m, the average particle is so close to the bottom of the bucket that we encounter the situation where $\psi_{\rm{min}} \approx \psi_{\rm{max}}$. We are then forced to choose $\psi_{\rm{targ}} > \psi_{\rm{max}}$, thus placing the average particle outside the bucket. This marks the beginning of the out-of-bucket regime.

Even though the average particle is outside the bucket, a fraction of the particles in the microbunch are still inside the bucket. For the next two periods of the oscillation, a part of the microbunch follows the in-bucket trajectories, while a part of the microbunch follows the out-of-bucket trajectories. The average particle, tracing the average behaviour of the entire microbunch, moves in and out of the bucket. At $z = 49$~m, $\psi_{\rm{max}} = 0$, indicating that the average particle is at the same energy level as the lowest point of the bucket.

In the out-of-bucket regime, $\psi_{\rm{min}}$ continues to increase, and becomes closer and closer to $\pi$. However, we want to prevent the average particle from getting too close to $\pi$, or a fraction of the particles in the microbunch will enter the acceleration region associated with the bucket ahead. This concern forces us to choose $\psi_{\rm{targ}} < \psi_{\rm{min}}$. The consequence is that the average particle enters the $\psi < 0$ region, thus absorbing energy from the radiation. Hence, $\bar{\eta}$ no longer decreases monotonically [see Fig.~\ref{Fig:evolution_curves}(b)]. Instead, $\bar{\eta}$ decreases and increases within a single undulator segment.

The final saturation regime begins at $z = 55.5$~m, where $\psi_{\rm{min}} = \pi$ [see Fig.~\ref{Fig:evolution_curves}(c)]. In this regime, it is no longer possible to prevent the average particle from entering the $\psi < 0$ region, regardless of the choice of $\psi_{\rm{targ}}$. As the average particle enters deep into the $\psi < 0$ region, the energy extraction becomes far less effective.

\subsection{Direct observation in phase space}

\begin{figure*}[p]
   \centering
   \includegraphics*[width=0.88\textwidth]{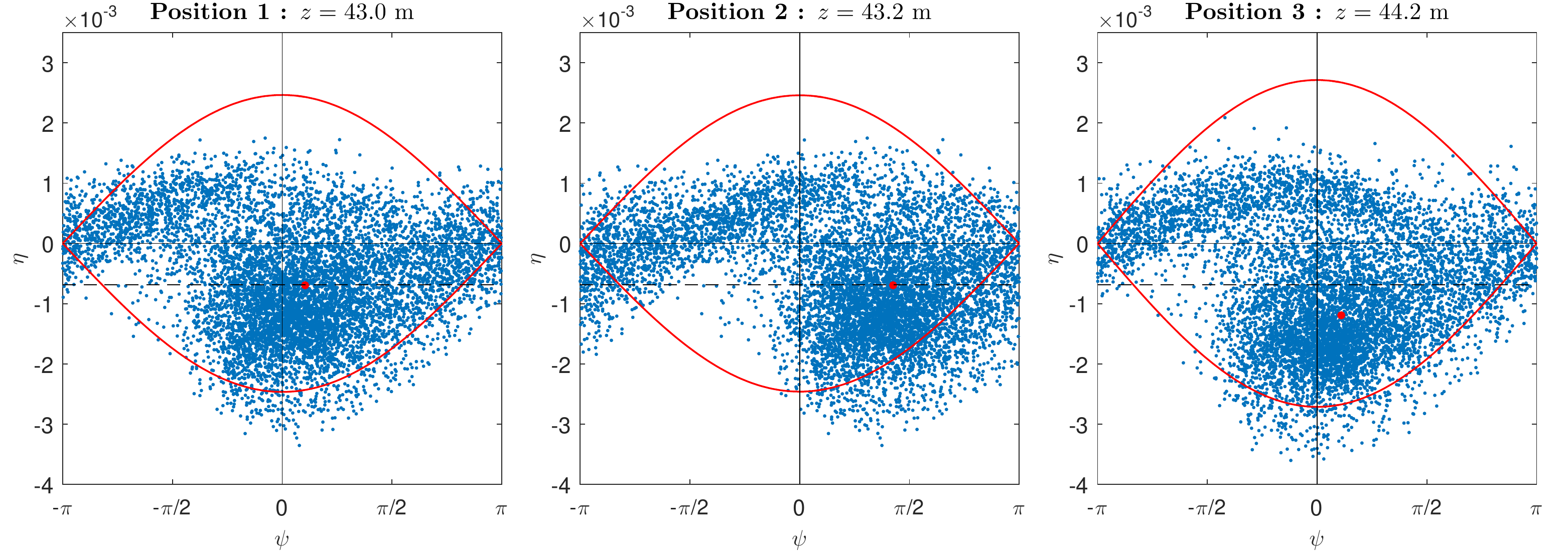}
   \caption{Simulation results. Snapshots in the longitudinal phase space $(\psi, \eta)$ showing a microbunch deceleration cycle in the in-bucket regime.}
   \label{Fig:comic_strip_1}
 \vspace{0.3cm}
   \includegraphics*[width=0.88\textwidth]{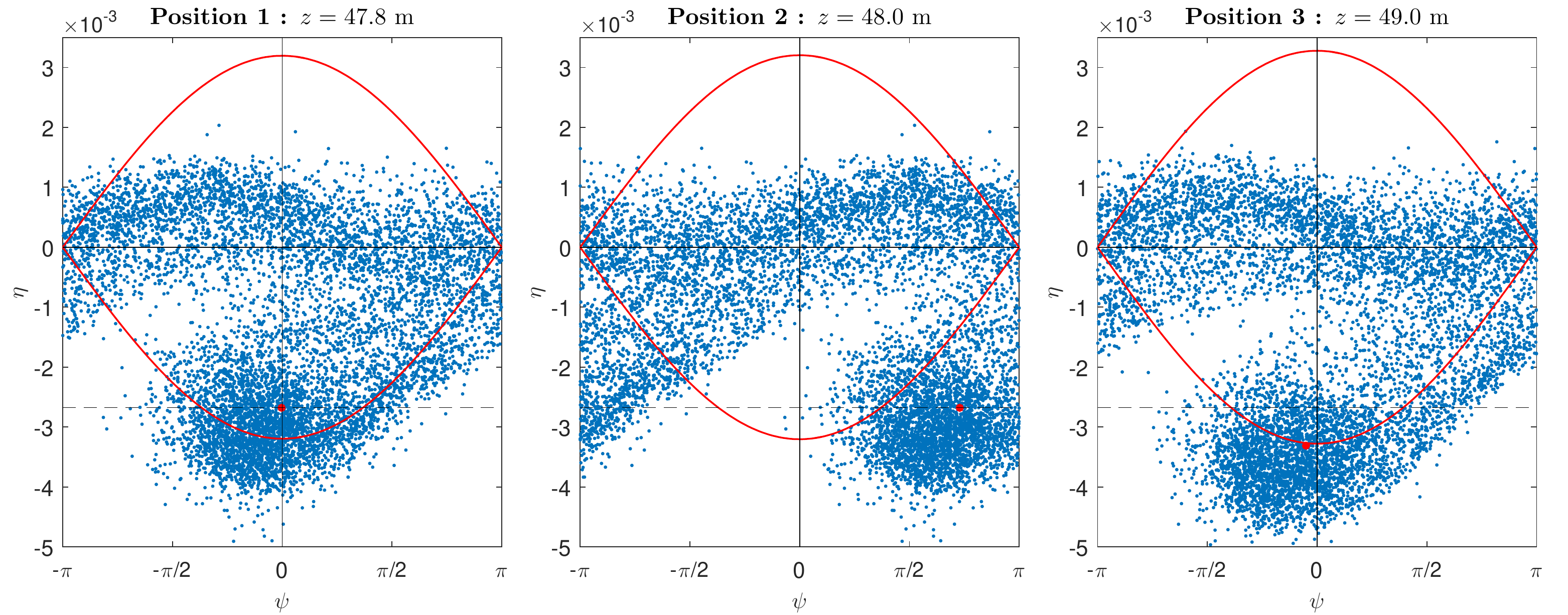}
   \caption{Simulation results. Snapshots in the longitudinal phase space $(\psi, \eta)$ showing a microbunch deceleration cycle during the transition from the in-bucket regime to the out-of-bucket regime.}
   \label{Fig:comic_strip_2}
 \vspace{0.3cm}
   \includegraphics*[width=0.88\textwidth]{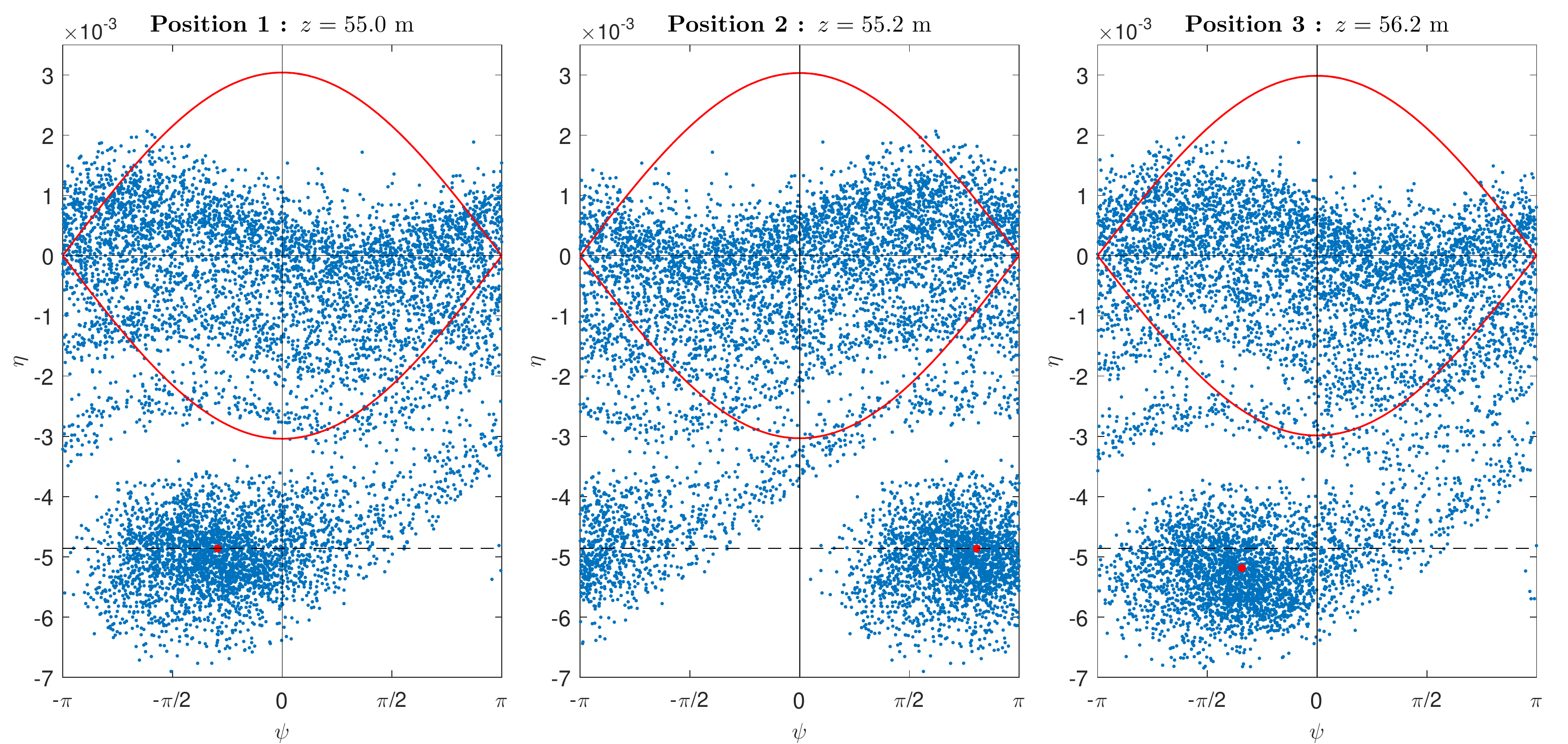}
   \caption{Simulation results. Snapshots in the longitudinal phase space $(\psi, \eta)$ showing a microbunch deceleration cycle at the onset of the final saturation regime.}
   \label{Fig:comic_strip_3}
\end{figure*}

Next, we observe the microbunch deceleration cycle directly in the longitudinal phase space $(\psi, \eta)$. Figure~\ref{Fig:comic_strip_1} shows a cycle in the in-bucket regime. Figure~\ref{Fig:comic_strip_2} shows a cycle during the transition from the in-bucket regime to the out-of-bucket regime. Figure~\ref{Fig:comic_strip_3} shows a cycle at the onset of the final saturation regime.

In each of these phase space snapshots, the red curve represents the separatrix of the ponderomotive bucket, and the red dot represents the average particle. These snapshots clearly show that the electron beam remains microbunched along the undulator line. From these snapshots, it is also apparent that the microbunch moves towards lower $\eta$ as $z$ increases, verifying the microbunch deceleration once again.

In each cycle, position~1 corresponds to the start point of the drift section, position~2 the end point of the drift section, and position~3 the end point of the subsequent undulator segment.

In position~1 and position~2, the average particle has the same $\eta$. This is expected, as the phase jump $\psi_{\rm{jump}}$ changes only the phase, but not the energy, of the average particle.

In the transition from position~2 to position~3, the particles pass through an undulator segment, where there is energy exchange between the particles and the radiation. The energy exchange alters the bucket half-height $h$ slightly. This is also expected, as $h$ depends on the slowly varying optical field amplitude $E_0$ according to the proportionality~(\ref{Eq:h_propto_E0}).

In all the three cycles shown here, the average particle ends up with a lower $\eta$ at position~3 than at position~1, meaning that there is a net energy transfer from the microbunch to the radiation in the undulator segment. The motion of the average particle in these snapshots reflects the mechanism depicted in Figs.~\ref{Fig:decel_cycle} and~\ref{Fig:out-of-bucket_mechan}.

\subsection{Trace of the average particle}

\begin{figure}[tb]
   \centering
   \includegraphics*[width=0.9\columnwidth]{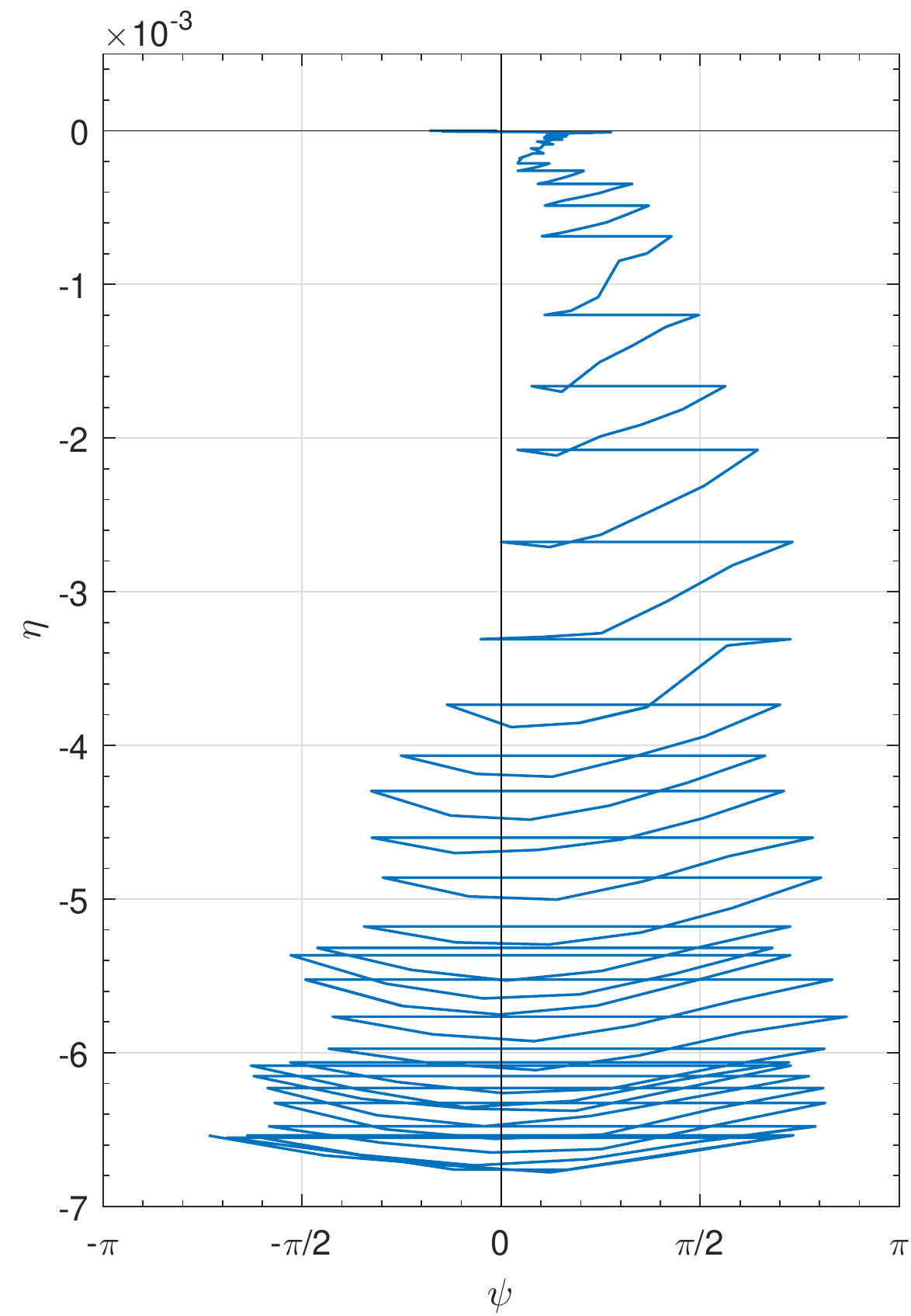}
   \caption{Simulation result. The trace of the average particle in the longitudinal phase space $(\psi, \eta)$ over the entire undulator line.}
   \label{Fig:avg_par_trajectory}
\end{figure}

Figure~\ref{Fig:avg_par_trajectory} shows the trace of the average particle in the longitudinal phase space $(\psi, \eta)$ over the entire undulator line. Within the trace, there are straight, horizontal sections, each representing the transition from position 1 to position 2 within a microbunch deceleration cycle.

As the average particle moves from high $\eta$ to lower $\eta$, the $\eta$ spacing between successive horizontal sections decreases, meaning that the amount of energy lost by the average particle decreases with every cycle. In other words, the deceleration efficiency decreases with $\eta$, as predicted by Eq.~(\ref{Eq:decel_eff}).

At $\eta \approx - 6 \times 10^{-3}$, the $\eta$ spacing between successive horizontal sections approaches zero. There is no longer a net energy transfer from the microbunch to the radiation. This indicates the arrival of the final saturation.

\subsection{Undulator segment length}

In the simulation, the synchrotron period $L_{\rm{sync}}$ varies as a function of $z$ (data not shown). The smallest value is $\min[L_{\rm{sync}}(z)] = 5.9$~m, which occurs at $z = 49$~m. With a undulator segment length of $L_{\rm{segm}} = 1$~m as specified in Table~\ref{Table:sim_par}, the requirement~(\ref{Eq:Lsegm_required}) is satisfied.

\section{Time-dependent effects}

Our physics model of the phase jump method is a steady-state one. For the purpose of verifying the model, we have conducted the numerical simulations in the steady-state mode.

So far, time-dependent effects have not been taken into consideration. To lay the foundation for future studies, we dedicate this section to a brief discussion on a time-dependent phenomenon, namely, the growth of sidebands in the FEL power spectrum.

For undulator tapering, sideband growth is a known issue, which degrades the spectral brightness at the desired FEL wavelength. This is discussed, for instance, in Refs.~\cite{KMR},~\cite{Quimby}, and~\cite{Claudio}.

The phase jump method has a similar nature to undulator tapering as a technique for enhancing the FEL power beyond the initial saturation. This prompts us to question whether sideband growth is an issue in the phase jump method as well.

The origin of the sidebands is the oscillation of the particle phase $\psi$ with the distance $z$ along the undulator line. For undulator tapering, the period of the oscillation is the synchrotron period $L_{\rm{sync}}$. This causes the growth of sidebands at~\cite{Quimby}
\begin{equation}
\frac{\Delta \lambda}{\lambda} = \pm \frac{\lambda_u}{L_{\rm{sync}}}.
\end{equation}

For the phase jump method, provided that the undulator segment length $L_{\rm{segm}}$ satisfies the requirement~(\ref{Eq:Lsegm_required}), the average particle never undergoes one complete orbit within the ponderomotive bucket after the initial saturation, as the orbit is disrupted by the applied phase jumps. Nonetheless, the phase $\bar{\psi}$ of the average particle still oscillates along the undulator line, as seen in Fig.~\ref{Fig:evolution_curves}(c). The period of this oscillation is the distance between successive phase shifters, given by $L_{\rm{segm}} + L_{\rm{drift}}$. Therefore, we expect this oscillation to trigger sidebands at
\begin{equation}
\label{Eq:sideband_wavelengths}
\frac{\Delta \lambda}{\lambda} = \pm \frac{\lambda_u}{L_{\rm{segm}} + L_{\rm{drift}}}.
\end{equation}

According to Eq.~(\ref{Eq:sideband_wavelengths}), it is, in principle, possible to influence the sidebands by varying $L_{\rm{segm}}$ and $L_{\rm{drift}}$ as functions of $z$.

With the parameter values listed in Table~\ref{Table:sim_par}, Eq.~(\ref{Eq:sideband_wavelengths}) yields $\Delta \lambda / \lambda = \pm 16 \times 10^{-3}$, which lie outside the FEL bandwidth. In attempt to verify this prediction, we repeat the numerical simulation in time-dependent mode. The resulting graph of radiation power versus $z$ (not shown here) agrees qualitatively with Fig.~\ref{Fig:evolution_curves}(a). However, sidebands are not seen in the average power spectrum of 30 shots (not shown here). Further investigation is needed to achieve a full understanding of time-dependent effects in the phase jump method.

\section{Conclusion}

In this article, we have examined the underlying physics of the phase jump method for enhancing the efficiency of an FEL. We have developed a one-dimensional, steady-state physics model, and we have verified it with three-dimensional, steady-state numerical simulations.

The physics model illustrates the post-saturation energy extraction process in the longitudinal phase space. It considers an average particle within the microbunch, and describes the microbunch deceleration cycle inside and outside the ponderomotive bucket. The model sets out the selection criteria for the target phase in each phase jump, describes the mechanism of the final saturation, and gives an upper limit for the undulator segment length. With the aid of our physics model, we have also discussed the similarities and differences between the phase jump method and undulator tapering.

In addition, we have given a brief discussion of time-dependent effects in the phase jump method, to lay the foundation for future studies.

\end{document}